%% file: model-arxiv.tex
\documentclass{article}
\pdfoutput=1  

\usepackage{cite} 
\usepackage{url}  
\usepackage{graphicx}

\newcommand{\eq}[1]{Eq.~(\ref{eq.#1})} 
\newcommand{\eqbare}[1]{(\ref{eq.#1})} 
\newcommand{\fig}[1]{Fig.~\ref{fig.#1}}

\newcommand{\tbl}[1]{Table~\ref{table.#1}}

\newcommand{\sectlabel}[1]{\label{sect.#1}}
\newcommand{\eqlabel}[1]{\label{eq.#1}}
\newcommand{\figlabel}[1]{\label{fig.#1}}
\newcommand{\tbllabel}[1]{\label{table.#1}}

\newcommand{\Pvisiblity}{P_{\mbox{\scriptsize visibility}}}
\newcommand{\popularity}{p_{\mbox{\scriptsize popularity}}} 
\newcommand{\Pother}{\beta} 

\newcommand{\fractionToPage}{f_{\mbox{\scriptsize page}}} 
\newcommand{\frontPageGrowth}{k_{\mbox{\scriptsize f}}} 
\newcommand{\newPageGrowth}{k_{\mbox{\scriptsize u}}} 

\newcommand{\Votes}{v}			
\newcommand{\submitterfanVotes}{v_S} 	
\newcommand{\fanVotes}{v_F} 	
\newcommand{\nonfanVotes}{v_N} 

\newcommand{\Users}{U}			
\newcommand{\submitterfans}{S}	
\newcommand{\fans}{F} 			
\newcommand{\nonfans}{N}		

\newcommand{\Plognormal}{P_{\mbox{\scriptsize lognormal}}}
\newcommand{\Pprior}{P_{\mbox{\scriptsize prior}}}

\newcommand{\rMax}{r_{\mbox{\scriptsize max}}} 

\newcommand{\probUserIsAFan}{\rho}

\newcommand{\Rsubmitterfan}{r_S} 	
\newcommand{\Rfan}{r_F} 	
\newcommand{\Rnonfan}{r_N} 

\newcommand{\Psubmitterfan}{P_S} 	
\newcommand{\Pfan}{P_F} 	
\newcommand{\Pnonfan}{P_N} 

\newcommand{\Csubmitterfan}{c_S} 	
\newcommand{\Cfan}{c_F} 	
\newcommand{\Cnonfan}{c_N} 

\newcommand{\Tpromotion}{T_{\mbox{\scriptsize promotion}}}

\newcommand{\erfc}{\mbox{erfc}} 

\newcommand{\hour}{\mbox{hr}}

\newcommand{\state}[1]{{\tt #1}}

\newcommand{\figwidth}{3.5in}
\newcommand{\figwidthWide}{5.5in}
\begin{document}

\title{Social Dynamics of Digg}

\author{Tad Hogg\\ Institute for Molecular Manufacturing \\
    Palo Alto, CA 94301        \and
        Kristina Lerman\\ USC Information Sciences Institute \\
    Marina del Rey, CA 90292
}

\maketitle

\begin{abstract}
Online social media provide multiple ways to find interesting content. One important method is highlighting content recommended by user's friends.
We examine this process on one such site, the news aggregator Digg. With a stochastic model of user behavior, we distinguish the effects of the content visibility and interestingness to users. We find a wide range of interest and distinguish stories primarily of interest to a users' friends from those of interest to the entire user community. We show how this model predicts a story's eventual popularity from users' early reactions to it, and estimate the prediction reliability. This modeling framework can help evaluate alternative design choices for displaying content on the site.

\end{abstract}

\section{Introduction}

The explosive growth of the Social Web hints at collective problem-solving made possible when people have tools to connect, create and organize information on a massive scale. The social news aggregator Digg, for example, allows people to collectively identify interesting news stories. The microblogging service Twitter has created a cottage industry of third-party applications, such as identifying trends from the millions of conversations taking place on the site and notifying you when your friends are nearby. Other sites enable people to collectively create encyclopedias, develop software, and invest in social causes. Analyzing records of complex social activity can help identify communities and important individuals within them,  
suggest relevant readings, 
and identify events 
and trends.

Effective use of this technology requires understanding how the social dynamics emerges from the decisions made by interconnected individuals. One approach is a stochastic modeling framework, which
represents each user as a stochastic process with a few states.  
As an example, applying this approach to Digg successfully described observed voting patterns of Digg's users~\cite{Lerman07ic,hogg09b,diggtist}.
However, quantitative evaluation of the model was limited by the poor quality of data, which was extracted by scraping Digg's web pages. 

In this paper we present two refinements to this modeling approach for Digg. First, we explicitly allow for systematic differences in \emph{interest} in news stories for linked and unlinked users. This distinction is a key aspect of social media where links indicate commonality of user interests. We also include additional aspects of the Digg user interface in the model, thereby accounting for cases where the existing model identified anomalous behaviors.
As the second major contribution, we describe how to measure confidence intervals of model predictions. We show that confidence intervals are highly correlated with the error between the predicted and actual votes stories accrue. Thus the confidence intervals indicate the quality of the model's predictions on a per-user or per-story basis.

This paper is organized as follows. The next section describes Digg and our data set. We then present a stochastic model of user behavior on Digg that explicitly includes dependencies on social network links. Using this model, we quantify these dependencies and discuss how the model predicts eventual popularity of newly submitted content. 
Finally, we compare our approach with other studies and discuss possible applications of stochastic models incorporating social network structure.

\section{Digg: A Social News Portal}
\sectlabel{digg}
At the time data was collected, Digg was a popular news portal with over 3 million registered users.
Digg allowed users to submit and rate news stories by voting on, or `digging', them. 
Every day Digg promoted a small fraction of submitted stories to the highly visible \emph{front page}.
Although the exact promotion mechanism was kept secret and changes occasionally, it appears to use the number of votes the story receives. Digg's popularity was largely due to the emergent front page created by the collective decisions of its many users. Below we describe the user interface that existed at the time of data collection.

\subsection{User interface}
Submitted stories appear in the \emph{upcoming} stories list, where they remain for about 24 hours or until promoted to the front page.
By default, Digg shows upcoming and front page stories in recency lists i.e., in reverse chronological order with the most recently submitted (promoted) story at the top of the list. A user may choose to display stories by popularity or by some broad topic. Popularity lists show stories with the most votes up to that time, e.g., the most popular stories submitted (promoted) in the past day or week. Each list is divided into pages, with 15 stories on each page, and the user has to click to see subsequent pages.

Digg allows users to designate friends and track their activities. The friend relationship is asymmetric. When user $A$ lists user $B$ as a \emph{friend}, $A$ can follow the activities of $B$ but not vice versa. We call $A$ the \emph{fan}, or follower, of $B$. The \emph{friends interface}  shows users the stories their friends recently submitted or voted for.\footnote{At the time of data collection Digg offered a social filtering feature which recommended stories, including upcoming stories, that were liked by users with a similar voting history. It is not clear how often users employed these features and we do not explicitly include them in our model.}

In this paper, we focus on the recency and ``popular in the last 24 hours'' lists for all stories and the friends interface list for each user. These lists appear to account for most of the votes a story receives.

\subsection{Evolution of story popularity}

\begin{figure}[tb]
\centering
\includegraphics[width=\figwidth]{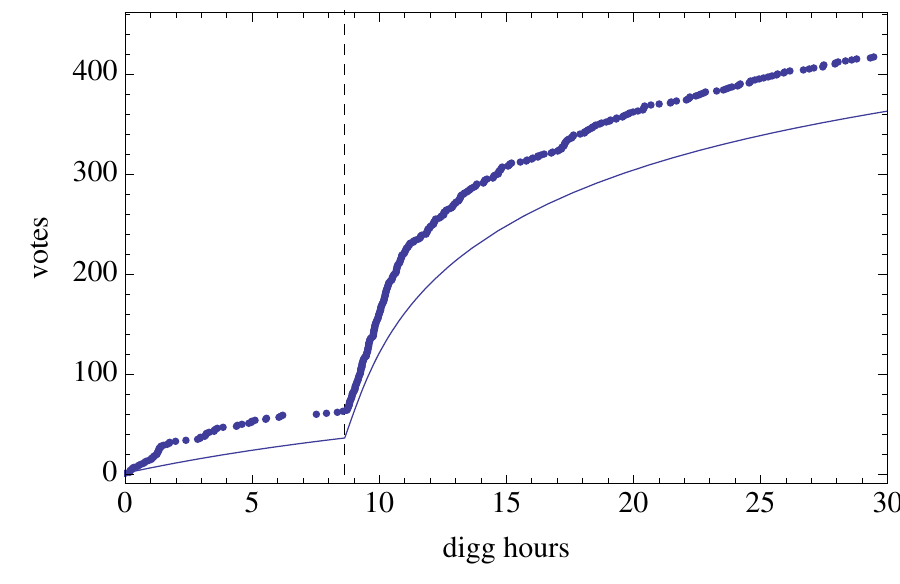}
\caption{Voting behavior: the number of votes vs.~time, measured in Digg hours, for a promoted story. The curve shows the corresponding solution from our model and the dashed vertical line indicates when the story was promoted to the front page. This story eventually received 452 votes.
}\figlabel{vote example}
\end{figure}

Most Digg users focus on front page stories, so upcoming stories accrue votes slowly. When a story is promoted to the front page, it becomes visible to many more users and accrues votes rapidly. \fig{vote example} shows the evolution of the number of votes for a story submitted in June 2009. The slope abruptly increases at promotion time (dashed line). As the story ages, accumulation of new votes slows down, and after a few days stories typically no longer receive new votes.

The final number of votes varies widely among the stories. Some promoted stories accumulate thousands of votes, while others muster only a few hundred.
Stories that are never promoted receive few votes, in many cases just a single vote from the submitter, and are removed after about 24 hours.

A challenge for understanding this variation in popularity is the interaction between the stories' \emph{visibility} (how Digg displays them) and their \emph{interestingness} to users. Models accounting for the structure of the Digg interface can help distinguish these contributions to story popularity.

\subsection{Data set}
We used Digg API to collect complete (as of July 2, 2009) voting histories of all stories promoted to the front page of Digg in June 2009.\footnote{The data set is available at http://www.isi.edu/$\sim$lerman/downloads/digg2009.html}
For each story, we collected story id, submitter's id, and the list of voters with the time of each vote. We also collected the time each story was promoted to the front page. The data set contains over 3 million votes on 3,553 promoted stories. We did not retrieve data about stories that were submitted to Digg during that time period but were never promoted. Thus our focus is on the behavior of promoted stories, which receive most of the attention from Digg users.

We define an \emph{active user} as any user who voted for at least one story on Digg during the data collection period. Of the 139,409 active users, 71,367 designated at least one other user as a friend. We extracted the friends of these users and reconstructed the fan network of active users, i.e., a directed graph of active users who are following activities of other users.

Over the period of a month, some of the voters in our sample deleted their accounts and were marked ``inactive'' by Digg. Such cases represent a tiny fraction of all users in the data set; therefore, we take the number of users to be constant.

\subsection{Daily activity variation}

\begin{figure}[t]
\centering
\includegraphics[width=\figwidth]{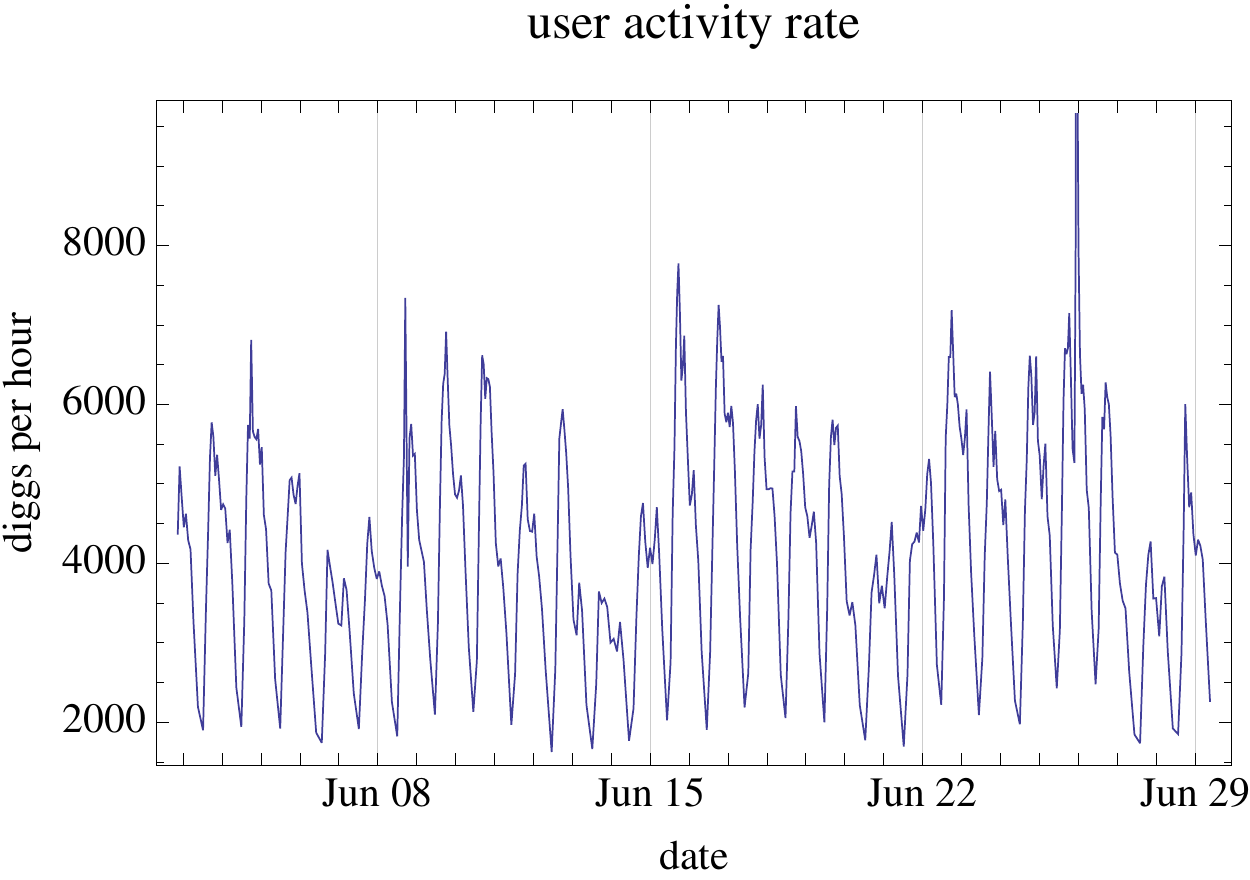}
\caption{Voting rate (diggs per hour) on front page stories during June 2009. The indicated dates are the start of each day (0:00 GMT). The minimum in daily activity is around noon GMT.} \figlabel{digg time}
\end{figure}

Activity on Digg varies considerably over the course of a day, as seen in \fig{digg time}. Adjusting times by the cumulative activity on the site accounts for this variation and improves predictions. Following~\cite{szabo09,hogg10b} we define the ``Digg time'' between two events (e.g., votes on a story) as the total number of votes made during the time between those events. With our data, this only counts votes on stories that were eventually promoted to the front page. In our data set, there are on average about 4000 votes on front page stories per hour, with a range of about a factor of 3 in this rate during the course of a day. This behavior is similar to that seen in an extensive study of front page activity in 2007~\cite{szabo09}, and as in that study we scale the measure by defining a ``Digg hour'' to be the average number of front page votes in an hour.

\section{Social Dynamics of Digg}

A key challenge in stochastic modeling is finding a useful combination of simplicity, accuracy and available data to calibrate the model. 
Stochastic models of online social media describe the joint behavior of their users and content. Since these web sites receive much more content than users have time or interest to examine, one important property to model is how readily users can find content. A second key property is how users react to content once they find it. Thus an important modeling choice for social media is the level of detail sufficient to distinguish user behavior and content visibility.
Following the practice of population dynamics~\cite{Haberman87} and epidemic modeling~\cite{Hethcote00} we consider groups of users and content. We assume that individuals within each group  have sufficiently similar behavior that their differences do not affect the main questions of interest. In the case of Digg, one such question is the number of votes a story receives over time. In our approach, we 
focus on how a single story accumulates votes, based on the combination of how easily users can find the story  and how interesting it is to different groups of users.

Following Ref.~\cite{Lerman07ic,hogg09b}, we start with a simple model in which story visibility is determined primarily by its location on the recency and friends lists, and use a single value to describe the story's interestingness to the user community. 
We use the ``law of surfing''~\cite{huberman98}  to relate location of the story to how readily users find it. This model successfully captured the qualitative behavior of typical stories on Digg and how that behavior depended on the number of fans of the story's submitter~\cite{hogg09b,Lerman10www}.

However, the simple model did not quantitatively account for several behaviors in the new data set. These included the significant daily variation in activity rates seen in \fig{digg time} and systematic differences in behavior between fans of a story's submitter and  other users. In particular, the new data was sufficiently detailed to show users tend to find stories their friends submit as more interesting than stories friends vote on but did not submit.
Another issue with the earlier model is a fairly large number of votes on stories far down the recency list. This is especially relevant for upcoming stories where the large rate of new submissions means a given story remains near the top of the recency list for only a few minutes. To account for such votes, the model's estimated ``law of surfing'' parameters indicated users browse an implausibly large number of pages while visiting Digg.

These observations motivate the more elaborate model described in this paper. This model includes systematic differences in interestingness between fans and other users and additional ways Digg makes stories visible to users.

\begin{figure}[tb]
\centering
\fbox{
\includegraphics[viewport=75 50 750 570,clip,width=\figwidth]{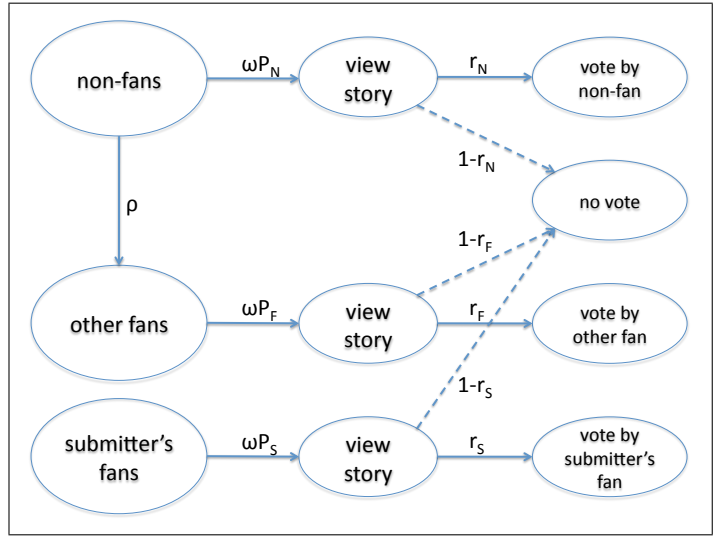}
}
\caption{State diagram for a user with respect to a particular story.}\figlabel{state diagram}
\end{figure}

\begin{figure}[tb]
\centering
\fbox{
\includegraphics[viewport=60 370 585 520,clip,width=\figwidth]{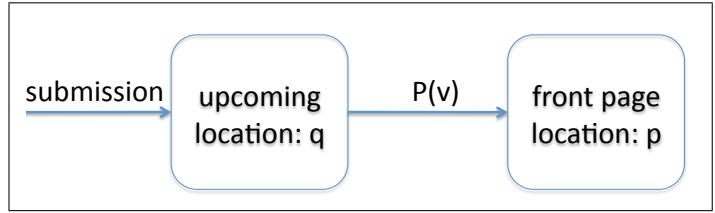}
}
\caption{State diagram for a story. A story not promoted after sufficient time (usually within a day) is removed (a state transition not shown in the diagram).
}\figlabel{state diagram story}
\end{figure}

\subsection{User model}
We allow for differences between users by separating them into groups, and assume that visibility of stories and voting behavior of users within each group is statistically the same. 
We refine the previous stochastic model of Digg~\cite{hogg09b} by not only distinguishing votes from fans and non-fans~\cite{hogg10b}, but also allowing for differences between fans of the \emph{submitter} and fans of \emph{other} voters who are not also fans of the submitter. The  state diagram \fig{state diagram} shows the resulting user model. The state \state{submitter's fans} includes all users who are fans of the submitter and have not yet seen the story; the \state{other fans} state includes all users who are fans of other voters but not the submitter, who have not yet seen the story; and the \state{non-fans} state includes all users who are neither fans of the submitter nor other voters, and have not yet seen the story. The state \state{no vote} includes all users who have seen the story and decided not to vote for it.
With respect to votes on a given story treated in this model, users visit Digg according to a Poisson process with average rate $\omega$ in terms of Digg time.

Users transition between states stochastically by browsing Digg's web pages and voting for stories.
The submitter provides a story's first vote. All of her fans start in the \state{submitter's fans} state, and all other users start in the \state{non-fans} state.  Each vote causes non-fan users who are that voter's fans and who have not yet seen the story to transition from the \state{non-fans} state to the \state{other fans} state. A user making this transition is not aware of that change until later visiting Digg and seeing the story on her friends list.

Once a user sees a story, she will vote for it with probability given by how {interesting} she finds the story.
Nominally people become fans of those whose contributions they consider interesting, suggesting fans have a systematically higher interest in stories than non-fans. Our model accounts for this by having the probability a user votes on a story depend on the user's state.
Users in each state also have a different probability to see stories, which is determined by the story's \emph{visibility} to that category of users.
Users vote at most once on a story, and our focus is on the final decision to vote or not after the user sees the story.

The visibility of stories changes as stories age and accrue votes. \fig{state diagram story} shows the state diagram of stories. A story starts at the top of the upcoming stories recency list.
The location increases with each new submission.
A promoted story starts at the top of the front pages.
The location increases as additional stories are promoted.

These state diagrams lead to a description of the average rates of growth~\cite{Lerman11tist} for votes from submitter fans, other fans and non-fans of prior voters, $\submitterfanVotes$, $\fanVotes$ and $\nonfanVotes$, respectively:
\begin{eqnarray}
\frac{d \submitterfanVotes}{dt} &=& \omega \Rsubmitterfan \Psubmitterfan \submitterfans \eqlabel{vS} \\
\frac{d \fanVotes}{dt} &=& \omega \Rfan \Pfan \fans \eqlabel{vF}\\
\frac{d \nonfanVotes}{dt} &=& \omega \Rnonfan \Pnonfan \nonfans \eqlabel{vN}
\end{eqnarray}
where $t$ is the Digg time since the story's submission and $\omega$ is the average rate a user visits Digg (measured as a rate per unit Digg time). 
We find only a small correlation between voting activity and the number of fans. Thus we use the average rate users visit Digg, rather than having the rate depend on the number of fans a user has.
$\nonfanVotes$ includes the vote by the story's submitter.  $\Psubmitterfan$, $\Pfan$ and $\Pnonfan$ denote the story's \emph{visibility} and $\Rsubmitterfan$, $\Rfan$ and $\Rnonfan$ denote the story's \emph{interestingness} to users who are submitter fans, other fans or non-fans of prior voters, respectively. Visibility depends on the story's state (e.g., whether it has been promoted), as discussed below. Interestingness is the probability a user who sees the story will vote on it.

These voting rates depend on the number of users in each category who have not yet seen the story:  $\submitterfans$, $\fans$ and $\nonfans$. The quantities change as users see and vote on the story, with average rate of change given by
\begin{eqnarray}\eqlabel{fans}
\frac{d \submitterfans}{dt} &=& -\omega \Psubmitterfan \submitterfans \eqlabel{S}\\
\frac{d \fans}{dt} &=& -\omega \Pfan \fans + \probUserIsAFan \nonfans \frac{d \Votes}{dt} \eqlabel{F}\\
\eqlabel{nonfans}
\frac{d \nonfans}{dt} &=& -\omega \Pnonfan \nonfans - \probUserIsAFan \nonfans \frac{d \Votes}{dt} \eqlabel{N}
\end{eqnarray}
with $\Votes =\submitterfanVotes+ \fanVotes + \nonfanVotes$ the total number of votes the story has received. The quantity $\probUserIsAFan$ is the probability a user is a fan of the most recent voter, conditioned on that user not having seen the story nor being a fan of any voter prior to the most recent voter. For simplicity, we treat this probability as a constant over the voters, thus averaging over the variation due to clustering in the social network and the number of fans a user has.
The first term in each of these equations is the rate the users see the story. The second terms arise from the rate the story becomes visible in the friends interface of users who are not fans of previous voters but are fans of the most recent voter.

Initially, the story has one vote (from the submitter) and the submitter has $S_0$ fans, so $\submitterfanVotes(0) = \fanVotes(0)=0$, $\nonfanVotes(0)=1$, $\submitterfans(0)=S_0$, $\fans(0)=0$ and $\nonfans(0)=\Users-S_0-1$ where $\Users$ is the total number of active users at the time the story is submitted. Over time, a story becomes less visible to users as it moves down the upcoming or (if promoted) front page recency lists, thereby attracting fewer votes and hence fewer new fans of prior voters. If the story gathers more votes than other stories, it is moved to the top of the popularity list, so becomes more visible.

\begin{figure}[t]
\centering
\includegraphics[width=\figwidthWide]{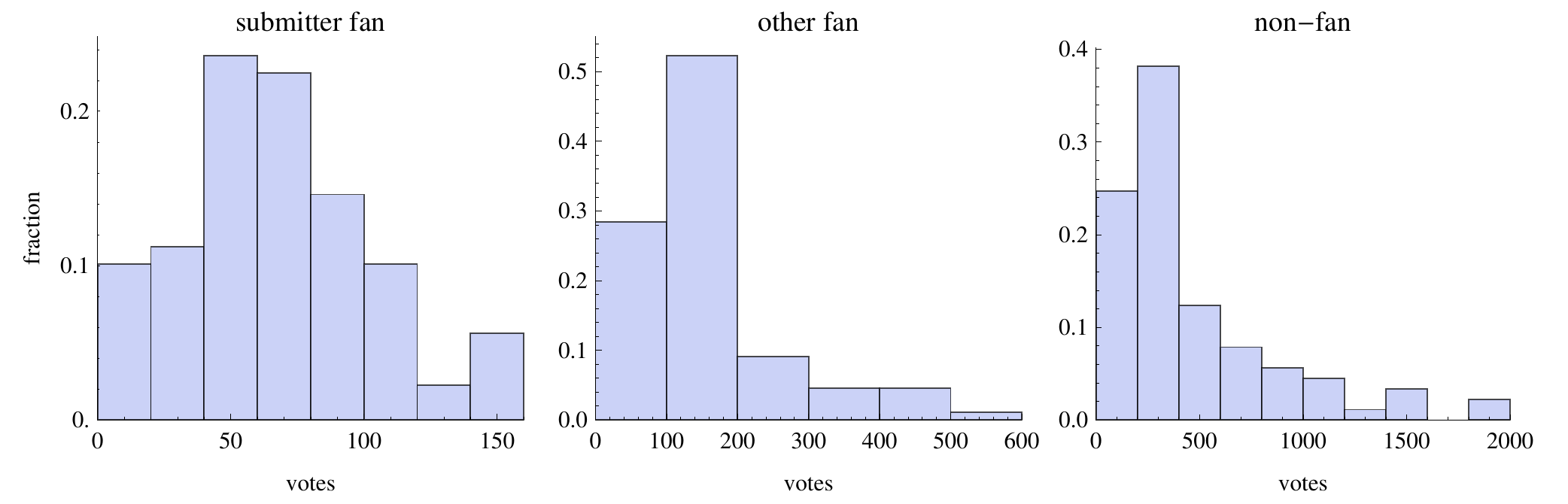}
\caption{Distribution of each type of vote a story accumulates 24 hours after promotion.}\figlabel{vote distributions}
\end{figure}

\fig{vote distributions} shows the range of votes the stories receive by 24 hours after promotion. Generally, stories have most votes from non-fans, somewhat fewer from other fans and a relatively small number from submitter's fans. The number of votes from submitter's fans is weakly correlated with the numbers from other fans (correlation coefficient $0.09$) and non-fans ($0.05$). The numbers from other fans and non-fans are highly correlated ($0.90$).

\subsection{Story visibility}

A fan easily sees the story via the friends interface, so we take $\Psubmitterfan=\Pfan=1$ for front page stories. While the story is upcoming, it appears in the friends interface but users do not necessarily choose to view upcoming stories friends liked.
Users can readily make this choice because the friends interface distinguishes upcoming from front page stories.
 We characterize the lower visibility of upcoming stories with constants $\Csubmitterfan$ and $\Cfan$ which are less than 1. The corresponding visibility is then $\Psubmitterfan=\Csubmitterfan$ and $\Pfan=\Cfan$.

Users who are not fans of prior voters must find the story on the front or upcoming pages. Thus $\Pnonfan$ depends on how users navigate through these pages and the story's location at the time the user visits Digg. This navigation is not given by our data. Instead, we use a model of user navigation through a series of web pages that has users estimating the value of continuing at the
site, and leaving when that value becomes
negative~\cite{huberman98}. This ``law of surfing'' leads to an inverse Gaussian
distribution of the number of pages $m$ a user visits before leaving
the web site,
\begin{equation}\eqlabel{stopping distribution}
e^{-\frac{\lambda  (m-\mu )^2}{2 m \mu ^2}} \sqrt{\frac{\lambda}{2 \pi m^3}}
\end{equation}
with mean $\mu$ and variance $\mu^3/\lambda$.
We use this distribution for user navigation on Digg~\cite{hogg09b}.

The visibility of a story on the $m^{th}$ front or upcoming page is the fraction of users who visit \emph{at least} $m$ pages, i.e., the upper cumulative distribution of \eq{stopping distribution}. For $m>1$, this fraction is
\begin{equation}
\fractionToPage(m) = \frac{1}{2}\left( F_m(-\mu) - e^{2\lambda/\mu} F_m(\mu) \right)
\end{equation}
where $F_m(x)=\erfc(\alpha_m (m-1+x)/\mu)$, $\erfc$ is the complementary error function, and $\alpha_m = \sqrt{\lambda/(2(m-1))}$. For $m=1$, $\fractionToPage(1)=1$.
The visibility of stories decreases in two distinct ways when a new story arrives. First, a story moves down the list on its current page. Second, a story at the $15^{th}$ position moves to the top of the next page. For simplicity, we model these processes as decreasing visibility in the same way through $m$ taking on fractional values within a page, e.g., $m=1.5$ denotes the position of a story half way down the list on the first page.

Digg presents several lists of stories. We focus on two lists as the major determinants of visibility for front page stories: reverse chronological order (``recency'') and most popular in the past 24 hours (``popularity''). Users can also find stories via other means. For instance, Digg includes other lists showing recent and popular stories in specific topics (e.g., sports or business) and popularity over longer time periods, e.g., the previous week. Stories on Digg may also be linked to from external web sites (e.g., the submitter's blog).

For front page votes, the recency and popularity lists provide the bulk of non-fan votes while the stories are close to the top of at least one of these lists, as illustrated in \fig{front page vote locations}. The rank on the recency list is the number of stories promoted more recently than that story 
and the rank on the popularity list is the number of stories, promoted within the 24 hours prior to the vote, with more votes. Since a page shows 15 stories, the location in terms of number of pages, as shown in the figure, is $1/15^{th}$ the rank, starting from page 1. Some votes occur while stories are far down both the recency and popularity lists, so the user likely finds the story by another method.

\begin{figure}[tb]
\centering
\includegraphics[width=4in]{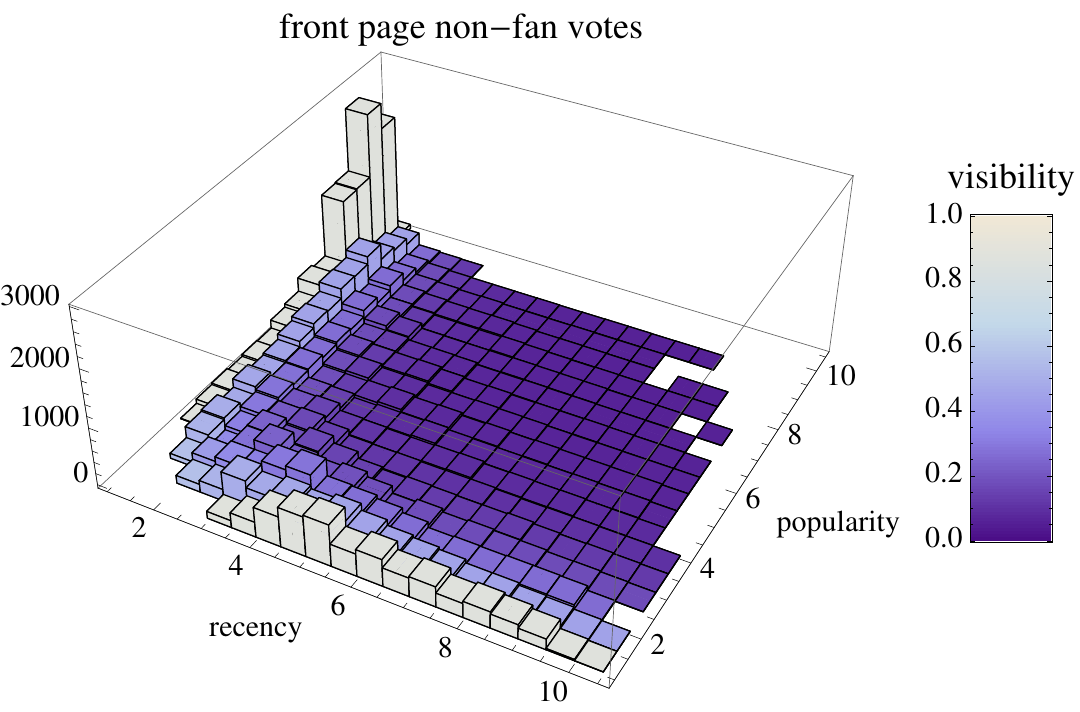}
\caption{Distribution of front page non-fan votes by location of the story on recency and popularity lists (for votes within 24 hours of story promotion), for a sample of 100 stories with a total of 41615 such votes. The colors indicate the visibility for each location predicted by the model parameters using \eq{visibility}, ranging between 0 and 1 as indicated on the legend.
}\figlabel{front page vote locations}
\end{figure}

From these observations, we account for three ways users who are not connected to prior voters find a story: via the recency list, via the popularity list or via one of the other methods described above. We combine visibility from these three methods assuming independent choices by users,  giving the probability to see the story as
\begin{equation}\eqlabel{visibility}
\Pvisiblity(t,v) = 1 -
( 1 - \fractionToPage(p(t)) ) \,  ( 1 - \fractionToPage(\popularity(v)) )\,  (1-\Pother)
\end{equation}
where $p(t)$ and  $\popularity(v)$  are the locations of the story on the recency and popularity lists, respectively, and $\Pother$ is the probability to find the story by another method. Although the positions of the stories on these lists depend on the specific stories submitted or promoted shortly after the story, these locations are approximately determined by the time $t$ and number of votes $v$ the story has, as described below. For visibility by other methods, we simply take $\Pother$ to be a constant, independent of story properties such as time since submission or number of votes. That is, we do not explicitly model factors affecting the visibility of stories by other methods, as the recency and popularity lists account for the bulk of the non-fan votes determined by our parameter estimates discussed below.

The location of a story on the recency and popularity lists could be additional state variables, which change as new stories are added and gain votes. Instead of modeling this in detail, we approximate these locations using the close relation between location and time (for recency) or votes (for popularity).

\paragraph{Position of a story in the recency list}
Using Digg time to account for the daily variation in activity on the site, the rate of story submission and promotion is close to linear. 
Thus the page number of a story on either the upcoming or front page recency list is~\cite{hogg09b}
\begin{equation}
p(t) = \left\{
	\begin{array}{ll}
	 \newPageGrowth t + 1	& \mbox{if $t<\Tpromotion$}\\
	\frontPageGrowth (t-\Tpromotion)+1 & \mbox{otherwise}
	\end{array}
\right.
\end{equation}
where $\Tpromotion$ is the time the story is promoted to the front page
and the slopes are given in
\tbl{parameters}. Since each page holds 15 stories, these
rates are $1/15^{th}$ the story submission and promotion rates,
respectively.

\paragraph{Position of a story in the popularity list}
The position of a story on the popularity list is the number of stories submitted or promoted in the previous 24 hours with more votes, for stories on upcoming or front page lists, respectively. The distribution of votes among stories in a 24 hour period is similar from day to day. Thus a story's position on the popularity list, determined by the location of its number of votes in this distribution, is approximately a function of its number of votes alone, with only minor variation depending on the time (i.e., the set of other stories from the past 24 hours). Thus we model the position as depending only on the number of votes the story has at the time, i.e., consider $\popularity(v)$ as a function of the number of votes the story has. This gives a simple, approximate relation between the actual location and number of votes -- ignoring the minor variations due to the specific stories promoted at different times.

\begin{figure}[tb]
\centering
\includegraphics[width=\figwidth]{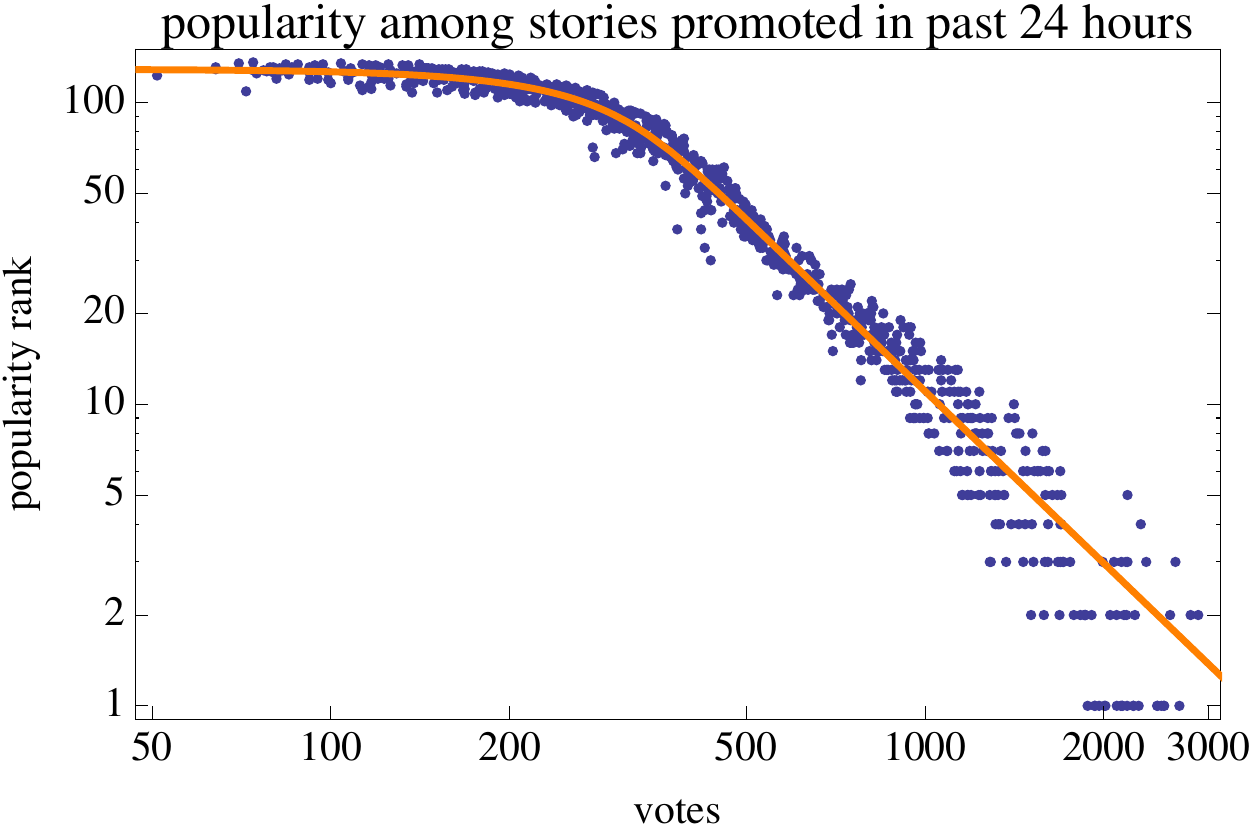}
\caption{Relation between rank on popularity list and number of votes for front page stories on a log-log plot. The curve is the fit to a double-Pareto lognormal distribution.}\figlabel{popularity location}
\end{figure}

\fig{popularity location} shows the relation between popularity rank and number of votes for a sample of front page votes within 24-hours of story promotion. To identify a suitable functional approximation for this relation, we note that a story typically accumulates votes at a rate proportional to how interesting it is to the user population. As seen in prior analysis of votes in 2006 on Digg~\cite{hogg09b}, we expect the interestingness to be lognormally distributed. Thus if we observe a sample of votes on stories over the same time interval for each story, the distribution of votes, and hence location on the popularity list, would follow a lognormal distribution. However, the popularity list includes stories of various times up to 24 hours since submission or promotion. Thus some stories of high interest will have few votes because they were just recently submitted or promoted, and conversely some stories with only moderate interestingness will have relatively many votes because they have been available for votes for nearly 24 hours. The combination of lognormal distribution of rates for accumulating votes and this variation in the observation times modifies the tails of the lognormal to be power-law, i.e., a double-Pareto lognormal distribution~\cite{reed04}.

Such a distribution fits the observed positions on the front page popularity list, as indicated in \fig{popularity location}.
The fit for the rank (i.e., number of stories above the given one in the popularity list, so the story promoted in the past 24 hours with the most votes has rank 0) is
\begin{equation}
\mbox{rank} = S (1-\Lambda(a,b,\nu,\sigma; v))
\end{equation}
where $S=129.0\pm0.1$ is the average number of stories promoted in 24 hours and $\Lambda(\ldots;v)$ is the cumulative distribution of a double-Pareto lognormal distribution, i.e., fraction of cases with fewer than $v$ votes. The parameters $a=1.90\pm0.005$ and $b=2.50\pm0.03$ are the power-law exponents for the upper and lower tails of the distribution, respectively, and the parameters $\nu=5.88\pm0.002$ and $\sigma=0.16\pm0.004$ characterize the location and spread of the lognormal behavior in the center of the distribution.
This fit captures the power-law tail relating stories near the top of the popularity list with the number of votes the story has. These are the cases for which the popularity list contributes significantly to the overall visibility of a story. More precisely, the Kolmogorov-Smirnov (KS) statistic shows the vote counts are consistent with this
distribution ($p$-value $0.92$).
We use this distribution, combined with the rate stories are promoted, to relate the number of votes a story has to its position on the popularity list, providing a functional form for $\popularity(v)$.

The popularity rank for upcoming stories submitted in the past 24 hours is more complicated than for the front pages due to the promotion. Stories with many votes are more likely to be promoted, and hence removed from the popularity list for upcoming stories. This removal alters the upper tail of the distribution and hence the numbers of votes for stories appearing near the top of the popularity list. Moreover, out of the $\approx 20,000$ stories submitted each day, our data includes only the $\approx 100$ stories per day eventually promoted. 
However, popularity significantly contributes to visibility only for stories near the top of the popularity list. Thus for our model, it is sufficient to determine the relation between votes and rank for upcoming stories with relatively many votes.
Such stories are likely to be promoted eventually and hence included in our sample. Instead of a power-law tail, our data on the eventually promoted stories is better fit by an exponential for the upcoming stories with relatively many votes, and hence near the top of the popularity list:
\begin{equation}
\mbox{rank} = e^{c - d v}
\end{equation}
with $c=5.3\pm0.01$ and $d=0.029\pm0.0002$.
This fits well for upcoming stories submitted within the past 24 hours with more than 100 votes, corresponding to rank of about 20 or less on the upcoming popularity list. For stories with few votes, e.g., fewer than 10 or 20, this fit based on the stories eventually promoted substantially underestimates the rank. Nevertheless, the estimated rank for such stories is still sufficiently large that  the law of surfing parameters we estimate indicate users do not find such stories via the popularity list. Thus this underestimate does not significantly affect our model's behavior for upcoming stories.

\paragraph{Friends interface}

The fans of the story's submitter can find the story via the friends
interface. As additional people vote on the story, their fans can
also see the story. We model this with $\fans(t)$, the number of fans of
voters on the story by time $t$ who are not also fans of the submitter and have not yet seen the story.
Although the number of fans is highly variable, we use the average number
of additional fans from an extra vote, $\probUserIsAFan \nonfans$, in \eq{fans}.

\section{Parameter estimation}

We estimate model parameters using 100 stories from the middle of our sample.

\subsection{Estimating parameters from observed votes}
In our model, story location affects visibility only for non-fan voters since fans of prior voters see the story via the friends interface. Thus we use just the non-fan votes to estimate visibility parameters, via maximum likelihood. Specifically, we use the non-fan votes to estimate the ``law of surfing'' parameters $\mu$ and $\lambda$, as well as the probability for finding the story some other way, $\Pother$. 

This estimation involves comparing the observed votes to the voting rate from the model. As described above, the model uses rate equations to determine the average behavior of the number of votes. We relate this average to the observed number of votes by assuming the votes from non-fan users form a Poisson process whose expected value is $d \nonfanVotes(t)/dt$, given by  \eq{vN}. 

For a Poisson process with a constant rate $v$, the probability to observe $n$ events in time $T$ is the Poisson distribution $e^{-v T} (v T)^n/n!$. This probability depends only on the \emph{number} of events, not the specific times at which they occur. Estimating $v$ involves maximizing this expression, giving $v = n/T$. Thus the maximum-likelihood estimate of the rate for a constant Poisson process is the average rate of the observed events.

In our case, the voting rate changes with time, requiring a generalization of this estimation. Specifically consider a Poisson process with nonnegative rate $v(t)$ which depends on one or more parameters to be estimated. Thus in a small time interval $(t,t+\Delta t)$, the probability for a vote is $v(t) \Delta t$, and this is independent of votes in other time intervals, by the definition of a Poisson process. Suppose we observe $n$ votes at times $0< t_1 < t_2, \ldots < t_n < T$ during an observation time interval $(0,T)$. Considering small time intervals $\Delta t$ around each observation, the probability of this observation is
\begin{eqnarray*}
P(\mbox{no vote in $(0,t_1)$}) \; v(t_1)\Delta t  & \times \\
P(\mbox{no vote in $(t_1,t_2)$}) \; v(t_2)\Delta t  &\times \\
 \ldots & \\
 P(\mbox{no vote in $(t_{n-1},t_n)$}) \; v(t_n)\Delta t  &\times \\
  P(\mbox{no vote in $(t_n,T)$})
\end{eqnarray*}
The probability for no vote in the interval $(a,b)$ is
\begin{displaymath}
\exp \left( -\int_a^b v(t) dt \right)
\end{displaymath}
Thus the log-likelihood for the observed sequence of votes is
\begin{equation}\eqlabel{log-likelihood}
 -\int_0^T v(t) dt  + \sum_i \log v(t_i)
\end{equation}
The maximum-likelihood estimation for parameters determining the rate $v(t)$ is a trade-off between these two terms:  minimizing $v(t)$ over the range $(0,T)$ to increase the first term while maximizing the values $v(t_i)$ at the times of the observed votes. If $v(t)$ is constant, this likelihood expression simplifies to $-v T + n \log v$ with maximum at $v=n/T$ as discussed above for the constant Poisson process. When $v(t)$ varies with time, the maximization selects parameters giving relatively larger $v(t)$ values where the observed votes are clustered in time.

We combine this log-likelihood expression from the votes on several stories, and maximize the combined expression with respect to the story-independent parameters of the model, while determining the interestingness parameters separately for each story.

\subsection{User activity} 

Our model involves a population of active users who visit Digg during our sample period and vote on stories.
Specifically, the model uses the rate users visit Digg, $\omega \Users$. We do not observe visits in our data, but can infer the relevant number of active users, $\Users$, from the heterogeneity in the number of votes by users. The data set consists of 139,409 users who voted at least once during the sample period, giving a total of 3,018,197 votes. \fig{user activity} shows the distribution of this activity. Most users have little activity during the sample period, suggesting a large fraction of users vote infrequently enough to never have voted during the time of our data sample. This behavior can be characterized by an activity rate for each user. A user with activity rate $\nu$ will, on average, vote on $\nu T$ stories during a sample time $T$.  We model the observed votes as arising from a Poisson process whose expected value is $\nu T$ and the heterogeneity arising from a lognormal distribution of user activity rates~\cite{hogg09c}:
\begin{equation}\eqlabel{lognormal}
\Plognormal(\mu,\sigma;r) = \frac{1}{\sqrt{2\pi}\, r \sigma} \exp \left(  -\frac{(\mu-\log(r))^2}{2\sigma^2} \right)
\end{equation}
where parameters $\mu$ and $\sigma$ are the mean and standard deviation of $\log(r)$.

\begin{figure}[tb]
\centering
\includegraphics[width=\figwidth]{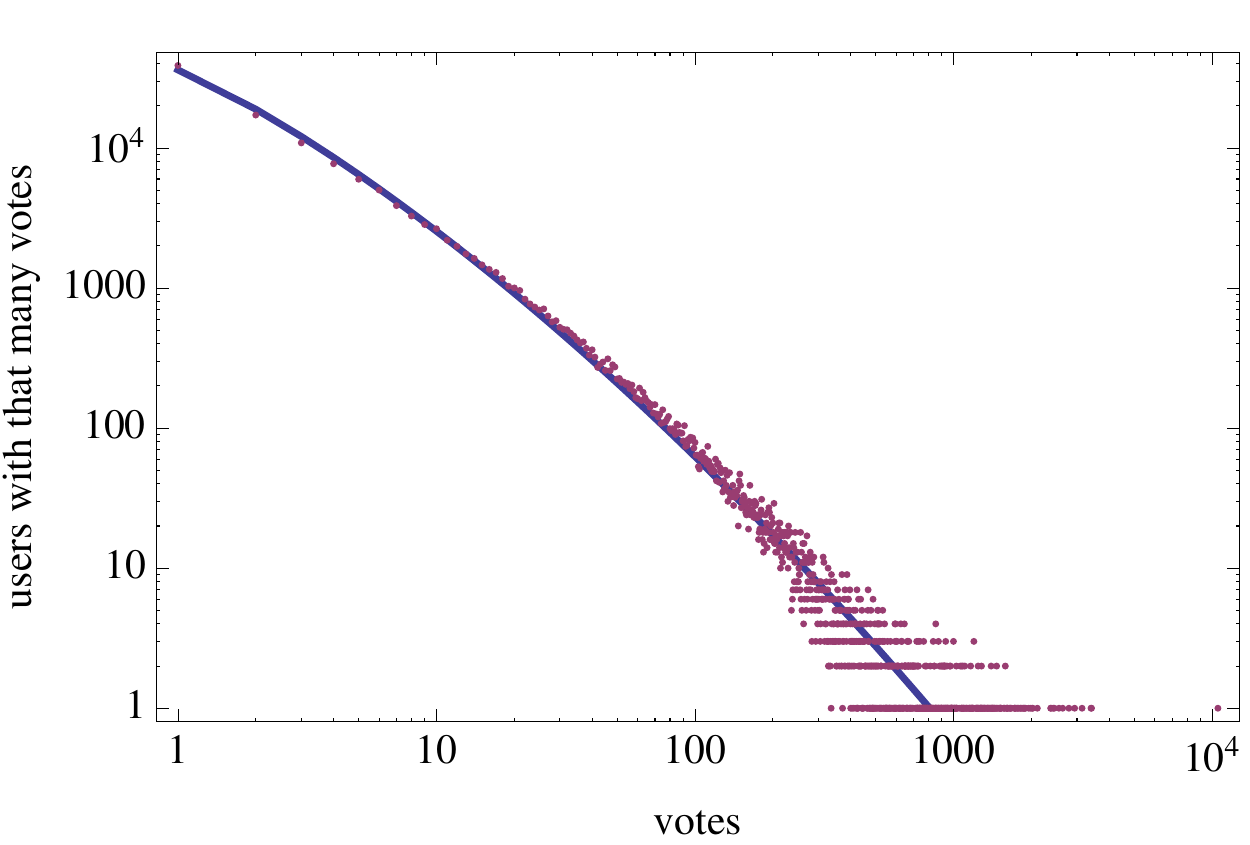}
\caption{User activity distribution on logarithmic scales. The curve shows the fit to the model described in the text.}\figlabel{user activity}
\end{figure}

This model gives rise to the extended activity distribution while accounting for the discrete nature of the observations. The latter is important for the majority of users, who have low activity rates and vote only a few times, or not at all, during our sample period.

Specifically, for $n_k$ users with $k$ votes during the sample period, this mixture of lognormal and Poisson distributions~\cite{bulmer74, miller07} gives the log-likelihood of the observations as
\begin{displaymath}
\sum_k n_k \log P(\mu,\sigma;k)
\end{displaymath}
where $P(\mu,\sigma;k)$ is the probability of a Poisson distribution to give $k$ votes when its mean is chosen from a lognormal distribution $\Plognormal$ with parameters $\mu$ and $\sigma$. From \eq{lognormal},
\begin{displaymath}
P(\mu,\sigma;k) = \frac{1}{{\sqrt{2 \pi }\, \sigma  k!}} \int_0^\infty   \rho ^{k-1} e^{-\frac{(\log (\rho )-\mu )^2}{2 \sigma ^2}-\rho
   } d\rho
\end{displaymath}
for integer $k\geq 0$. We evaluate this integral numerically. In terms of our model parameters, the value of $\mu$ in this distribution equals $\nu T$.

Since we do not observe the number of users who did not vote during our sample period, i.e., the value of $n_0$, we cannot maximize this log-likelihood expression directly. Instead, we use a zero-truncated maximum likelihood estimate~\cite{hilbe08} to determine the parameters $\mu$ and $\sigma$ for the vote distribution of \fig{user activity}. Specifically, the fit is to the probability of observing $k$ votes conditioned on observing at least one vote. This conditional distribution is $P(\mu,\sigma;k)/(1-P(\mu,\sigma;0))$ for $k>0$, and the corresponding log-likelihood is
\begin{displaymath}
\sum_{k>0} n_k \log P(\mu,\sigma;k) - U_{+} \log(1-P(\mu,\sigma;0))
\end{displaymath}
where $U_{+}$ is the number of users with at least one vote in our sample.
Maximizing this expression with respect to the distribution's parameters $\mu$ and $\sigma$ gives
$\nu T$ lognormally distributed with the mean and standard deviation of $\log(\nu T)$ equal to $-0.10\pm0.04$ and $2.43\pm0.02$, respectively. Based on this fit, the curve in \fig{user activity} shows the  expected number of users with each number of votes. This is a discrete distribution: the lines in the figure between the expected values serve only to distinguish the model fit from the points showing the observed values.

With these estimated parameters, $P(\mu,\sigma; 0)=0.43$, indicating $43\%$ of the users had sufficiently low, but nonzero, activity rate that they did not vote during the sample period. We use this value to estimate $\Users$, the number of active users during our sample period: $U = U_{+}/(1-P(\mu,\sigma; 0))$.

\subsection{Links among users}

We observe $u=258,218$ users with fans, and these users have a total of $c=1,731,658$ connections. Our data has 139,409 distinct voters, of which 78,007 have no fans. There is little correlation between links and voting activity, so we estimate the fraction of users with zero fans from the ratio of these values, i.e., about 56\%. Thus the average number of fans per user, including users without fans, is $c/(1.56 u) \approx 4.3$.

We estimate the model parameter $\probUserIsAFan$ of \eq{fans} and \eqbare{nonfans} as the probability a fan link connects the first to the second user of a randomly selected pair of users, corresponding to the average number of fans per user divided by the number of active users $\Users$.

\subsection{Visibility to submitter's fans} 

Because stories are always visible to fans and we know the number of fans of the story's submitter, the model behavior (\eq{vS} and \eqbare{S}) can be solved without reference to the rest of the model. We have $\Psubmitterfan=1$ when the story is on the front page and $\Psubmitterfan=\Csubmitterfan<1$, reflecting users' preference for front page stories. Thus, these equations have two story-independent parameters, i.e., the rate users visit Digg ($\omega$) and the probability users view upcoming stories submitted by their friends ($\Csubmitterfan$), and two story-dependent parameters, i.e., the interestingness ($\Rsubmitterfan$) and number of fans of the submitter ($S_0$).  $S_0$ is given in our data, while we estimate the other parameters from the data, i.e., votes by fans of the stories' submitters.

\subsection{Visibility to non-fans} 

In our model, story location affects visibility only for non-fan voters since fans of prior voters see the story via the friends interface. Thus we use just the non-fan votes to estimate visibility parameters, via maximum likelihood. A story typically receives only a few dozen votes before promotion, mostly from fans. With the value of $\probUserIsAFan$, estimated as described above, \eq{N} gives $\nonfans(t) \approx \Users$ up to a few hours after promotion. Over this time period, \eq{vN} simplifies to $d \nonfans/dt \approx \omega \Users \Rnonfan \Pnonfan$ with $\Pnonfan$ depending on story location on the recency and popularity lists. $\Rnonfan$ is constant for a given story, so $\Pnonfan$ determines the time variation in the voting rate by non-fans.

For front page stories, in our model $\Pnonfan=\Pvisiblity(t,v)$ from \eq{visibility}, which has three parameters: $\mu$ and $\lambda$ characterizing the browsing behavior for the recency and popularity lists, and the probability to find the story by other methods, $\Pother$. We estimate these parameters by maximizing the likelihood of observing the non-fan front page votes according to the model, as described above for estimating a Poisson process with a time-dependent rate in \eq{log-likelihood}. This estimation also determines $\Rnonfan$ for each story.

For upcoming stories, we take $\Pnonfan=\Cnonfan \Pvisiblity(t,v)$, giving a single additional parameter, $\Cnonfan$, to estimate, since we assume browsing behavior on the upcoming pages is the same as for front pages. This assumption has little effect on the model behavior because of the large number of submissions and relatively few non-fan votes for upcoming stories. A submitted story remains near the front of the recency list for only about a minute after submission and stories reaching the front of the popularity list (due to having many votes) are soon promoted to the front page. Thus moderate variations in how deeply users browse the upcoming recency or popularity lists (i.e., the values for $\mu$ and $\lambda$) have little effect on the non-fan votes. Instead, the relatively few non-fan upcoming votes arise mainly through users finding the story by other means.
That is, in most cases $\Pvisiblity \approx \Pother$ for upcoming stories. Thus $\Pnonfan=\Cnonfan \Pvisiblity(t,v) \approx \Cnonfan \Pother$ and any difference between $\Pother$ for upcoming and front page stories would merely rescale the value of $\Cnonfan$. This parameter is readily estimated using \eq{log-likelihood} with the upcoming non-fan votes.

\subsection{Visibility to other fans} 

From \eq{vF}, $d\fanVotes /dt$ changes abruptly when the story is promoted since $\Pfan$ changes from $\Cfan$ to 1 upon promotion. Thus we estimate $\Cfan$ by the change in voting rate by fans other than those of the submitter by comparing the votes a story receives one hour before promotion and the votes received during the hour after promotion.

With all story-independent parameters estimated, we can then solve the full model for a story to determine $d\fanVotes /dt$ as a function of time. This gives the expected rate of other fan votes as a function of time. We determine $\Rfan$ for the story as the value maximizing the log-likelihood (\eq{log-likelihood}) for the other fan votes the story receives.

\subsection{Summary}

\begin{table}[tb]
\begin{center}
{\small 
\begin{tabular}{l|l}
\hline parameter & value \\
\hline average rate each user visits Digg & $\omega=0.16 \pm 0.01\,/\hour$ \\
number of active users & $\Users = 248,000 \pm 3000$  \\ 
page view distribution  & $\mu=0.92 \pm 0.04$ \\
				& $\lambda=0.9 \pm 0.1$ \\
visibility by other methods	 & $\Pother=0.05 \pm 0.01$ \\
probability a user is a voter's fan & $\probUserIsAFan=1.7 \times 10^{-5}$ \\
upcoming stories location  & $\newPageGrowth =59.8\,\mbox{pages}/\hour$ \\ %
front page location  & $\frontPageGrowth = 0.31\,\mbox{pages}/\hour$ \\ %
 \multicolumn{2}{c}{fraction viewing upcoming pages\rule{0pt}{10pt}}  \\
submitter fans  	& $\Csubmitterfan=0.57 \pm 0.03$ \\
other fans  	& $\Cfan=0.10 \pm 0.01$ \\
non-fans  		& $\Cnonfan=0.11 \pm 0.01$ \\
\hline \multicolumn{2}{c}{story specific parameters\rule{0pt}{10pt}} \\
interestingness to submitter fans    & $\Rsubmitterfan$ \\
interestingness to other fans    & $\Rfan$ \\
interestingness to non-fans    & $\Rnonfan$ \\
number of submitter's fans  & $S_0$ \\
promotion time	& $\Tpromotion$ \\ 
\end{tabular}
}
\end{center}
\caption{Model parameters, with times in Digg hours.}\tbllabel{parameters}
\end{table}

\tbl{parameters} lists the estimated parameters.
All of these parameters, except the three story interestingness parameters $\Rsubmitterfan$, $\Rfan$ and $\Rnonfan$, are either known (e.g., the number of submitter's fans) or estimated from data from a sample of stories and then used for all stories. The interestingness parameters are estimated individually for each story from its votes.

\section{Results}

\fig{vote example} compares the solution of the rate equations with the actual votes for one story. This illustrates that the model captures the main qualitative features of the vote dynamics: an abrupt jump in votes after promotion followed by a slowing of the voting rate.

\fig{front page vote locations} shows how visibility estimated by our model (indicated by color) compares with the distribution of front page votes. Many votes occur when the story is recently promoted (so near the top of the recency list) or has received many votes within 24 hours after promotion (so near the top of the popularity list). This is consistent with our model, which predicts higher visibility for stories in these positions on the lists.

\subsection{Interestingness for fans and non-fans}

We use the model to evaluate systematic differences in story interestingness between fans and non-fans.
The estimated $r$ values indicate the stories have a wide range of interestingness to users, as shown in \fig{r values}, along with fits to lognormal distributions. The figure shows $\Rnonfan$ values tend to be much smaller than the interestingness for fans, as also seen in an earlier study with a smaller data set from 2006~\cite{hogg10b}. The $r$-values are weakly correlated, with Spearman rank correlation between $\Rsubmitterfan$ and $\Rfan$ of $0.20$, between $\Rsubmitterfan$ and $\Rnonfan$ of $0.22$, and between $\Rfan$ and $\Rnonfan$ of $0.13$.
Moreover, there is a large range in the ratio of interestingness to fans and non-fans, suggesting stories with particularly large ratios are mainly of niche interest.

\tbl{r distribution parameters} summarizes the lognormal distribution parameters.
A bootstrap test~\cite{efron79} based on the Kolmogorov-Smirnov (KS) statistic shows the estimated $r$-values are consistent with this distribution ($p$-value $0.11$, $0.14$ and $0.05$ for the three cases). This test and the others reported in this paper account for the fact that we fit the distribution parameters to the data~\cite{clauset07}.

\begin{table}
\begin{center}
\begin{tabular}{lcc}
vote type 		& $\mu$ 	& $\sigma$ \\ \hline
submitter fan 	& $-3.5 \pm 0.2$	& $0.8 \pm 0.1$ \\
other fan 		& $-2.3 \pm 0.1$	& $0.3 \pm 0.1$ \\
non-fan 		& $-6.3 \pm 0.1$	& $0.6 \pm 0.1$ \\
\end{tabular}
\end{center}
\caption{Parameters for lognormal distribution of interestingness.
}\tbllabel{r distribution parameters}
\end{table}

\begin{figure}[tbp]
\centering
 \includegraphics[width=\figwidthWide]{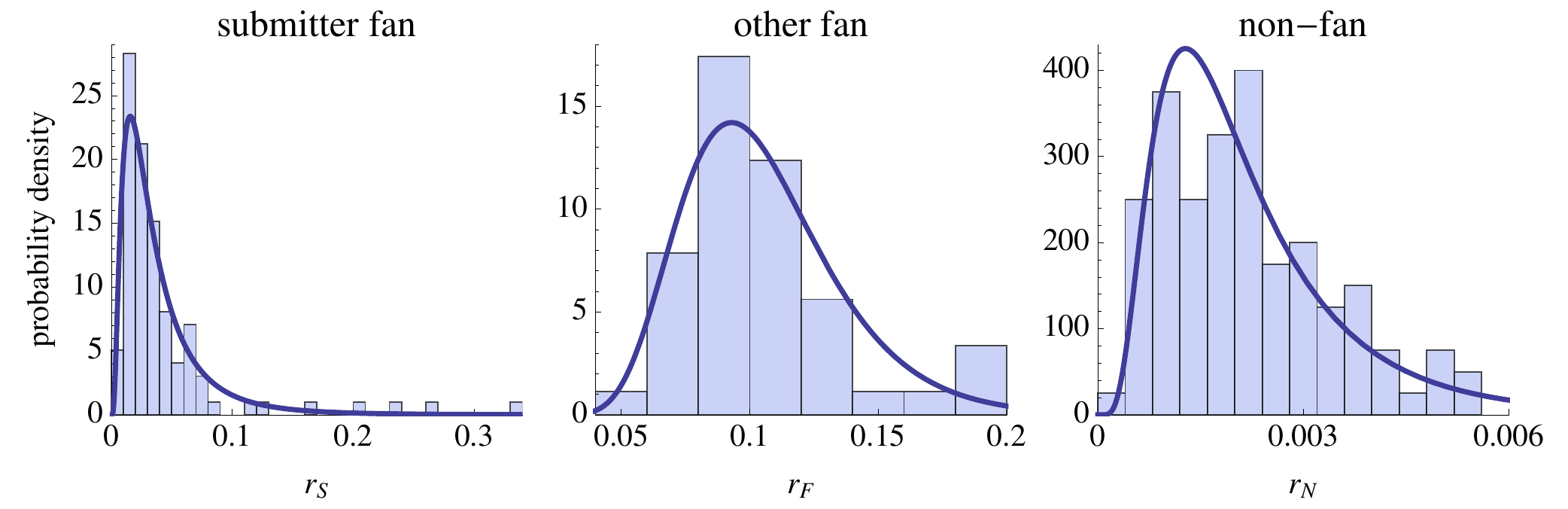}
\caption{Distribution of interestingness for each type of user.
The curves are  lognormal fits to the values. The plots have different axes scales.} \figlabel{r values}
\end{figure}

The relationship between interestingness for fans and other users indicates a considerable variation in how widely stories appeal to the general user community. Moreover, we find other fans have somewhat higher interest in stories than submitter fans,  i.e., $\Rfan$ tends to be larger than $\Rsubmitterfan$.
Since we have $\Csubmitterfan > \Cfan$ (\tbl{parameters}), we find submitter fans are more likely to view the story while upcoming, but less likely to vote for it, compared with other fans.
This suggests people favor the submitter as a source of stories to read, while the fact that a friend, not the submitter, voted for the story makes it more likely the user will vote for the story. 
Identifying these possibilities illustrates how models can suggest subgroups of behaviors in social media for future investigation.

\subsection{Predicting popularity from early votes}

In this section we use the model to predict popularity of Digg stories. We focus on the 89 of the 100 stories in the calibration data set that were promoted within 24 hours of submission. Most stories are promoted within 24 hours of submission (if they are ever promoted) and this restriction simplifies the model's use of the ``popular in last 24 hours'' list by not requiring it to check for removal from the list if the story is still upcoming more than 24 hours after submission.

Predicting popularity in social media from intrinsic properties of newly submitted content
is difficult~\cite{salganik06}. However, users' early reactions provide some measure of predictability~\cite{hogg09c,Kaltenbrunner07,Lerman08wosn,szabo09}.
The early votes on a story allow estimating its interestingness to fans and other users, thereby predicting how the story will accumulate additional votes. These predictions are for expected values and cannot account for the large variation due, for example, to a subsequent vote by a highly connected user which leads to a much larger number of votes.

We can improve predictions from early votes by using the lognormal distributions of $r$-values, shown in \fig{r values}, as the prior probability to combine with the likelihood from the observations according to Bayes theorem.
Specifically, instead of maximizing the likelihood of the observed votes, $P(r|\mbox{votes})$, as discussed above, this approach maximizes the posterior probability, which is proportional to $P(r|\mbox{votes}) \Pprior(r)$ where $\Pprior$ is taken to be the lognormal distribution $\Plognormal$ in \eq{lognormal} with parameters from the fits shown in \fig{r values}.

For a prediction at time $T$, we use the votes up to time $T$ to estimate the $r$ values by finding the values that maximize
\begin{equation}\eqlabel{likelihood with priors}
L = \log(P(\Rsubmitterfan,\Rfan,\Rnonfan|\mbox{votes})) + \log \left(\Pprior(\Rsubmitterfan) \Pprior(\Rfan) \Pprior(\Rnonfan) \right)
\end{equation}
We then solve the model starting at time $T$ and use the values from that solution as the predictions at later times. Solving the model equations starting at time $T$ requires initial values, i.e., the number of votes $\submitterfanVotes(T)$, $\fanVotes(T)$, $\nonfanVotes(T)$ and the size of the user groups who have not yet seen the story: $\submitterfans(T)$, $\fans(T)$, $\nonfans(T)$. The numbers of votes is available in our data. However, the sizes of the user groups is not available. Instead, we estimate these values from the voting \emph{rates} and the estimated $r$ values. For instance, \eq{vS} gives
\begin{equation}
\submitterfans(T) = \frac{1}{\omega \Rsubmitterfan \Psubmitterfan}  \frac{d \submitterfanVotes}{dt}
\end{equation}
We estimate the voting rates from the number of votes in the 15 minutes prior to time $T$, except if there are fewer than five votes in this time we extend the time interval to include the five previous votes. For simplicity, to avoid treating the discontinuity in visibility at promotion, we based this estimate on front page votes when $T$ is after the promotion time.

We focus on behavior after promotion. 
\fig{prediction example} compares predicted to actual votes for one story 24 hours after promotion. Votes from submission to promotion are used to estimate $r$ values for the three groups of users. The model solutions extend from the time of these estimates, i.e., the story's promotion time, to $t=24$ Digg hours after promotion. The model quantitatively reproduces the observed votes for this story.

\begin{figure}[tb]
\centering
\includegraphics[width=\figwidthWide]{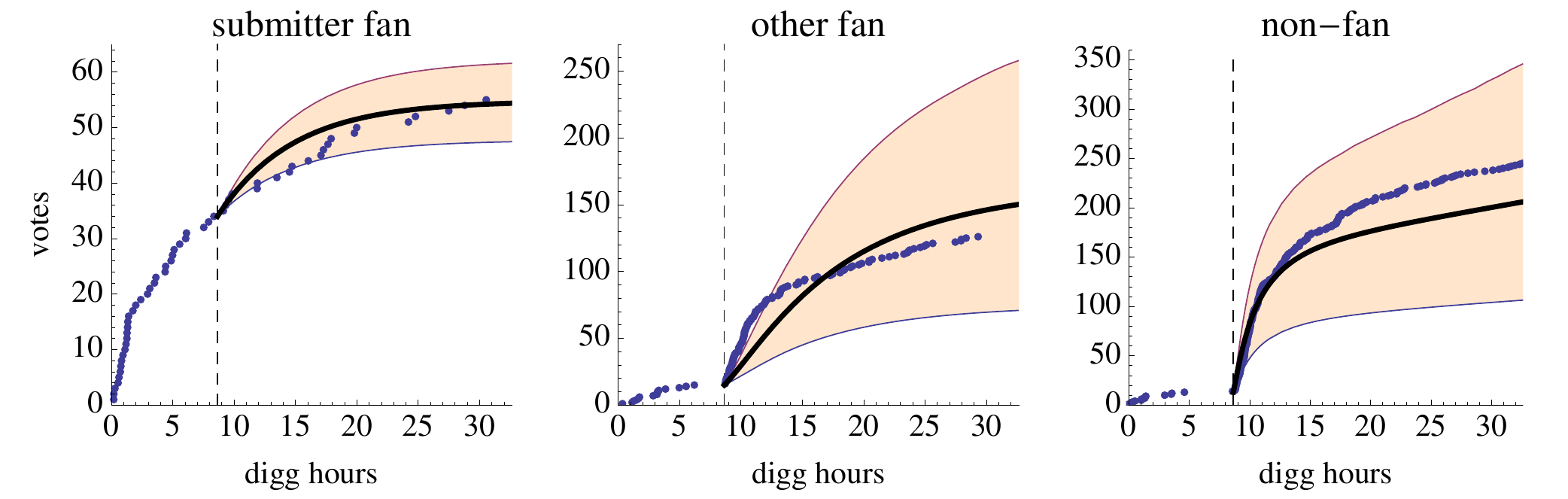}
\caption{Predictions compared to actual votes (dots) for each type of user for one story. The figure shows predictions made at promotion (black line) and the growth in the 95\% confidence interval of the prediction up to 24 hours after promotion. The dashed vertical line shows the story's promotion time.}\figlabel{prediction example}
\end{figure}

\begin{figure}[tb]
\centering
\includegraphics[width=\figwidth]{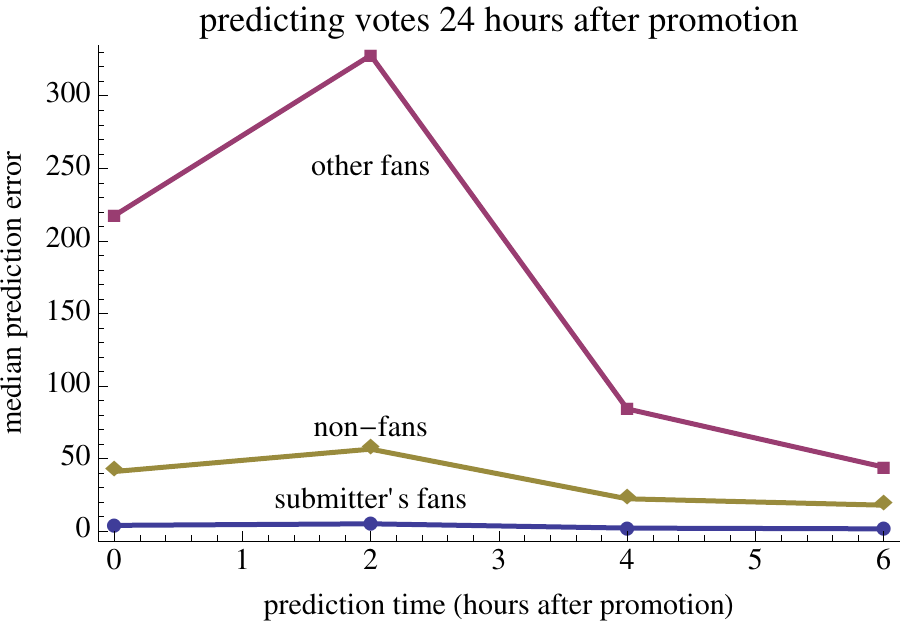}
\caption{Median error between predicted and observed votes 24 Digg hours after promotion for predictions made 0, 2, 4 and 6 Digg hours after promotion.}\figlabel{prediction}
\end{figure}

Generalizing from this example for a single story, \fig{prediction} shows the prediction errors for each type of vote the story receives 24 Digg hours after promotion, based on estimating $r$-values from early votes observed up to time $T$ for various times $T$ at and shortly after promotion. For context of the size of these errors, \fig{vote distributions} shows the range of number of votes the stories have at the times of these predictions, i.e., 24 hours after promotion.

As expected, errors generally decrease when predictions are made later. Of more interest is the difference among the type of votes, particularly for votes from other fans. Early votes are mainly from submitter's fans and non-fans, so the ability to predict differences in behavior for those groups based on early votes could be useful in quickly distinguishing stories likely to be of broad or niche interest to the user community.

Overall, the model reasonably predicts votes from submitter's fans and non-fans, but is much less accurate for votes from other fans. One reason for this difference is the relatively small number of other fan votes while a story is upcoming. Specifically, the pool of other fans $\fans$ starts at zero. Only a vote by a non-fan can increase $\fans$, and upcoming stories have low visibility to non-fan voters. Even after a pool of other fans becomes available, it takes some time for those users to return to Digg. Thus there are relatively few early other fan votes, leading to poor estimates for $\Rfan$ values. Moreover, the relatively small pool of other fans means a single early voter with many fans can significantly change $\fans$ away from its average value as used in the model. These factors lead to the relatively large errors in predicting the other fan votes. As a direction for future work, this observation suggests predictions would benefit from including measurements of the social network of the voters to determine the  value of $\fans$ at the time of prediction rather than using an estimate based on the model.

Another view of prediction quality is how well the model predicts the rank ordering of stories, i.e., whether the story is likely to be relatively popular. We measure this with the Spearman rank correlation between the model's prediction and the observed number of votes 24 Digg hours after promotion, as shown in \tbl{vote correlation}.  Even for other fan votes, where the absolute prediction error is relatively large, the predicted values give a good indication of the relative rank of the stories.

\begin{table}
\begin{center}
\begin{tabular}{cccc}
					& \multicolumn{3}{c}{\emph{correlation}} \\
$T-\Tpromotion$		& submitter fan 	& other fan & non-fan \\ \hline
0 		& $0.88$	& $0.26$ 	& $0.51$ \\
2 		& $0.95$	& $0.59$ 	& $0.85$ \\
4 		& $0.98$	& $0.69$ 	& $0.92$ \\
6 		& $0.99$	& $0.75$ 	& $0.94$ \\
\end{tabular}
\end{center}
\caption{Spearman rank correlation between predicted and observed number of each type of votes 24 Digg hours after promotion, for predictions made at various times $T$ after promotion (measured in Digg hours).
}\tbllabel{vote correlation}
\end{table}

Predicting whether a story will attract a large number of votes, rather than the precise number of votes, is a key issue for web sites such as Digg. Such predictions form the basis of using crowd sourcing to select a subset of submitted content to highlight~\cite{Lerman08wosn}.
As an example of this distinction, we predict whether a story will receive more than the median number of votes of each type of user based on votes received up to various times. This amounts to a binary classification task.
 \tbl{threshold prediction} compares predictions made at different times. The classification error rate is the fraction of stories for which prediction of whether the story receives more than the median number of votes differs from the actual value.

\begin{table}
\centering
\begin{tabular}{cccc}
					& \multicolumn{3}{c}{\emph{classification error}} \\
$T-\Tpromotion$		& submitter fan 	& other fan & non-fan \\ \hline
0 		& $0.19$	& $0.40$ 	& $0.29$ \\
2 		& $0.11$	& $0.48$ 	& $0.17$ \\
4 		& $0.10$	& $0.37$ 	& $0.15$ \\
6 		& $0.06$	& $0.28$ 	& $0.12$ \\
\end{tabular}
\caption{Classification errors on whether a story receives more than the median number of votes from each type of voter received by 24 Digg hours after promotion, for predictions made at various times $T$ after promotion (measured in Digg hours).}\tbllabel{threshold prediction}
\end{table}

\subsection{Confidence intervals}
We can use the model to estimate how well it predicts future votes. For a given set of parameter values, prediction variability comes from differences in estimated $r$ values. If $r$ is poorly determined, predictions will be unreliable.

To quantify this behavior, we numerically evaluate the second derivative matrix $D$ of the log-likelihood combined with the priors based on votes on the story up to time $T$, $L(r)$ given in \eq{likelihood with priors}, at the maximum $r=\rMax$, where $r=(\Rsubmitterfan,\Rfan,\Rnonfan)$. This gives
\begin{equation}
L(r) = L(\rMax) + \frac{1}{2} (r-\rMax) D (r-\rMax)
\end{equation}
to second order in $|r-\rMax|$. To this order of expansion, the likelihood is
\begin{equation}
\exp(-(r-\rMax) D (r-\rMax)/2)
\end{equation}
This corresponds to a multivariate normal distribution for $r$ with mean $\rMax$ and covariance matrix $-D^{-1}$. Since we are expanding around a maximum, the 2nd derivative matrix is negative definite so this gives a well-defined normal distribution, i.e., with a positive definite covariance matrix. This covariance includes both individual variances in the values of $\Rsubmitterfan$, $\Rfan$ and $\Rnonfan$ and correlations among their variations around the maximum.

If $L(r)$ is a fairly flat function of $r$ around the maximum, then maximum likelihood poorly constrains the values, corresponding to large variances in the normal distribution. Conversely, if $L(r)$ is sharply peaked, the distribution will be narrow.

We apply this observation to estimate confidence intervals for the predictions. We first numerically evaluate the second derivative matrix $D$ at the maximum. We then generate random samples of $r$ from the multivariate normal distribution. For each of these samples, we solve the model starting from the time $T$ to any desired time for predicting the votes, e.g., 24 hours after promotion. After collecting these predictions from many samples, we use quantiles of their ranges as the confidence intervals. In the examples presented here, we generate 1000 random samples and determine the 95\% confidence interval from the variation in $r$ values as the range between the $2.5\%$ and $97.5\%$ quantiles of these samples.

\begin{figure}[tb]
\centering
\includegraphics[width=\figwidthWide]{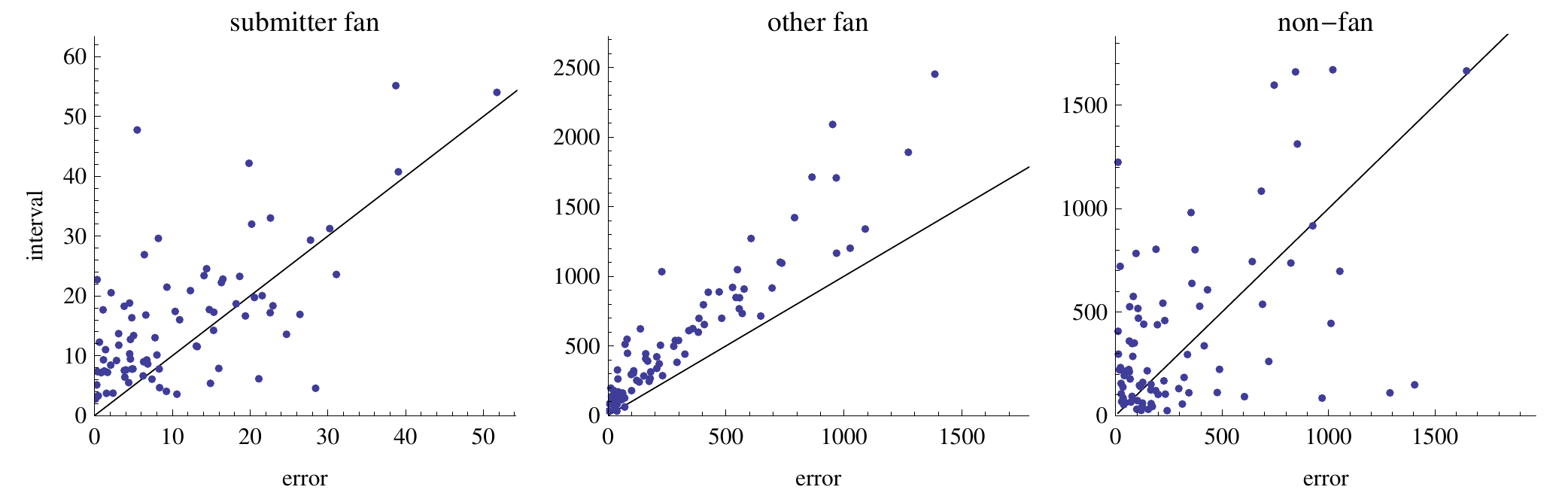}
\caption{Size of 95\% confidence interval vs.~prediction error 24 Digg hours after promotion for predictions based on votes up to each story's promotion time. The diagonal lines correspond to errors equal to the size of the confidence interval.}
\figlabel{prediction error and confidence}
\end{figure}

As one example, \fig{prediction example} shows how confidence intervals grow with time for predictions made from votes at the time a story is promoted. For multiple stories, \fig{prediction error and confidence} shows  the relation between 95\% confidence interval and prediction error 24 Digg hours after promotion, based on prediction made at the time of promotion. We see generally that large errors are associated with large confidence intervals, especially for the other fan votes where the model's prediction errors are largest. Thus the confidence intervals, which are computed from the vote information available at the time of prediction, indicate how well the model can predict votes.
These scatterplots show some cases where the error is considerably larger than the confidence interval.
They indicate additional sources of variation not accounted for by the variation in the estimated $r$ values. This could be due, for instance, to votes by exceptionally well-connected users that significantly increase the story's visibility compared to the average value assumed with the model.

\section{Related Work}
\sectlabel{related}

Models of social dynamics can help explain and predict the popularity of online content.
The broad distributions of popularity and user activity on many social media sites can arise from simple macroscopic dynamical rules~\cite{wilkinson08}. A phenomenological model of the collective attention on Digg describes the distribution of final votes for promoted stories through a decay of interest in news articles~\cite{wu07}. Stochastic models~\cite{Lerman07ic,hogg09b} offer an alternative explanation for the vote distribution. Rather than novelty decay, they explain the votes distribution by the combination of variation in the stories' inherent interest to users and effects of user interface, specifically decay in visibility as the story moves to subsequent pages.
Crane and Sornette~\cite{crane08} found that collective dynamics was linked to the inherent quality of videos on You\-Tube. From the number of votes received by videos over time, they could separate high quality videos
from junk videos. This study is similar in spirit to our own in exploiting the link between observed popularity and content quality. However, while these studies aggregated data from tens of thousands of individuals, our method focuses instead on the \emph{microscopic} dynamics, modeling how individual behavior contributes to content popularity.

Statistically significant correlation between early and late popularity of content is found on Slashdot~\cite{Kaltenbrunner07}, Digg and You\-Tube~\cite{szabo09}. Specifically, similar to our study, Szabo \& Huberman~\cite{szabo09} predicted long-term popularity of stories on Digg. Through large-scale statistical study of stories promoted to the front page, they were able to predict stories' popularity after 30 days based on its correlation with popularity one hour after promotion. Similarly, Lerman \& Hogg~\cite{Lerman10www} predicted popularity of stories based on their pre-promotion votes. We also quantitatively predict stories' future popularity, but unlike earlier works, we also estimate confidence intervals of these predictions for each story.

Previous works found social networks to be an important component to information diffusion. Niche interest content tends to spread mainly along social links in Second Life~\cite{bakshy09}, in blogspace~\cite{Colbaugh10isi}, as well as on Digg~\cite{Lerman08wosn}, and does not end up becoming very popular with the general audience.
Aral et al.~\cite{aral09} found that social links between like-minded people, rather than causal influence, explained much of  information diffusion observed on a network. Our modeling approach allows us to systematically distinguish users who are linked to those who are not linked and study diffusion separately for each group.

\section{Discussion}

Highlighting friends' contributions is a common feature of social media sites, including Digg.
To evaluate the effects of this behavior, we explicitly distinguish votes from submitter's fans, other fans and non-fans in our model, while separating the effects of differences in visibility and interestingness among these groups of users. This identifies that submitter's fans are, on average, far more likely to find the story interesting. Our model adjusts for the higher visibility of stories to fans, thereby identifying that increased attention from fans is not just due to the increased visibility. Identifying stories of particularly high interest to fans could be a useful guide for highlighting stories in the friends interface, i.e., emphasizing those with relatively large interestingness to friends as reflected in the early votes. Moreover, this information could be useful to recommend new fans to users, based on visibility-adjusted similarity in voting rather than, as commonly done in collaborative filtering~\cite{Konstan97grouplens}, just using the raw score of similar votes. This could be particularly important for users with relatively infrequent votes, where variations due to how visible a story is could significantly affect the similarity of the vote pattern with that of other users.

For more precise estimates, the web site could track the fraction of users seeing the story that vote for it, thereby directly estimating interestingness and accounting for the large variability in number of fans among the voters, in contrast to our model which used an average value. Exploiting such details of user behavior becomes more important as the complexity of the web site interface increases, offering many ways for users to locate content. Recording which method leads each user to find the story can aid in identifying any systematic differences in interests among those users. 

We find a wide range of interestingness ratios between fans and non-fans. This explains prior observations of the effect of relatively high votes from fans on indicating popularity to the general user population, and also suggests stories that are of niche interest to the fans rather than the general user population. Our assumption that fans of prior voters easily see the story is reasonable for users with relatively few friends, so only a few stories will appear in their friends interface. For users with many friends, visibility of a story would decrease when many newer stories appear on the friends interface. This possibility could be included in the model using the ``law of surfing'' for the stories appearing in each users' friends interface.

For prediction, we find the largest errors with votes from other fans. This likely arises from the relatively small number of such votes, especially while the story is upcoming. In that case, the large variation in number of fans per user can have a dramatic effect not accounted for in the model.
This suggests the main source of the prediction error arises from the long-tail distribution of fans per user, which the model treats as a single average value based on the parameter $ \probUserIsAFan$. We could test this possibility by collecting additional data on the actual fans of each voter, thereby using the observed value of $\fans(t)$ at the time of prediction when estimating $r$-values. In cases where $\fans(t)$ is particularly large, e.g., due to an early vote by a user with many fans, this will result in a smaller estimated value for $\Rfan$ and hence smaller predicted number of other fan votes.

Models can suggest improved designs for user-contributory web sites. Our results suggest it may be useful to keep popular stories visible longer for users who return to Digg less often -- giving them more chance to see the popular stories before they lose visibility. This would be a fine-tuned version of  ``popular stories'' pages, adjusted for each user's activity rate. That is, instead of showing stories in order of recency only, selectively move less popular stories down the page (once there are enough votes to determine popularity), thereby leaving the more popular ones nearer the top of the list for users who come back to Digg less often.

We examined behavior over a relatively short time (e.g., up to a day after promotion). Over longer times, additional factors could become significant, particularly a decrease in the interestingness as news stories submitted to Digg become ``old news''~\cite{wu07}.

Modeling visibility depends on how the web site user interface exposes content. This highlights a challenge for modeling social media: continual changes to the user interface can alter how visibility changes for newly submitted content.  Thus accurate models require not only data on user behavior but also sufficient details of the user interface at the time of the data to determine the relation between visibility and properties of the content.

The lognormal distribution of interestingness seen here and in other web sites~\cite{hogg09c} is useful as a prior distribution for estimating interestingness from early behavior on web sites. 
The use of such priors will be more important as models make finer distinctions among groups of users, e.g., distinguishing those who find the content in different ways as provided by more complex interfaces. In such cases, many groups will not be represented among the early reaction to new content and use of priors will be especially helpful.

User-contributory web sites typically allow users to designate others whose contributions they find interesting, and the sites highlight the activity of linked users. Thus our stochastic model, explicitly distinguishing behavior of users based on whether they are linked to users who submitted or previously rated the content, could apply to many such web sites.

\input{model-arxiv.bbl}

\end{document}

%% file: model-arxiv.bbl

\newcommand{\BMCxmlcomment}[1]{}

\BMCxmlcomment{

<refgrp>

<bibl id="B1">
  <title><p>Social Information Processing in Social News
  Aggregation</p></title>
  <aug>
    <au><snm>Lerman</snm><fnm>K.</fnm></au>
  </aug>
  <source>IEEE Internet Computing: special issue on Social Search</source>
  <pubdate>2007</pubdate>
  <volume>11</volume>
  <issue>6</issue>
  <fpage>16</fpage>
  <lpage>-28</lpage>
</bibl>

<bibl id="B2">
  <title><p>Stochastic Models of User-Contributory Web Sites</p></title>
  <aug>
    <au><snm>Hogg</snm><fnm>T</fnm></au>
    <au><snm>Lerman</snm><fnm>K</fnm></au>
  </aug>
  <source>Proc. of the Third International Conference on Weblogs and Social
  Media (ICWSM2009)</source>
  <publisher>AAAI</publisher>
  <pubdate>2009</pubdate>
  <fpage>50</fpage>
  <lpage>-57</lpage>
</bibl>

<bibl id="B3">
  <title><p>Using Stochastic Models to Describe and Predict Social Dynamics of
  Web Users</p></title>
  <aug>
    <au><snm>Lerman</snm><fnm>K</fnm></au>
    <au><snm>Hogg</snm><fnm>T</fnm></au>
  </aug>
  <source>To appear in ACM Transactions on Intelligent Systems and
  Technology</source>
  <pubdate>2012</pubdate>
</bibl>

<bibl id="B4">
  <title><p>Predicting the Popularity of Online Content</p></title>
  <aug>
    <au><snm>Szabo</snm><fnm>G</fnm></au>
    <au><snm>Huberman</snm><fnm>BA</fnm></au>
  </aug>
  <source>Communications of the ACM</source>
  <publisher>SSRN</publisher>
  <pubdate>2010</pubdate>
  <volume>53</volume>
  <issue>8</issue>
  <fpage>80</fpage>
  <lpage>-88</lpage>
</bibl>

<bibl id="B5">
  <title><p>Social Dynamics of {Digg}</p></title>
  <aug>
    <au><snm>Hogg</snm><fnm>T</fnm></au>
    <au><snm>Lerman</snm><fnm>K</fnm></au>
  </aug>
  <source>Proc. of the Fourth International Conference on Weblogs and Social
  Media (ICWSM2010)</source>
  <publisher>Menlo Park, CA: AAAI</publisher>
  <pubdate>2010</pubdate>
  <fpage>247</fpage>
  <lpage>-250</lpage>
</bibl>

<bibl id="B6">
  <title><p>Mathematical Models: Mechanical Vibrations, Population Dynamics,
  and Traffic Flow</p></title>
  <aug>
    <au><snm>Haberman</snm><fnm>R</fnm></au>
  </aug>
  <source>Paperback</source>
  <publisher>Society for Industrial Mathematics</publisher>
  <series><title><p>Classics in Applied Mathematics</p></title></series>
  <pubdate>1987</pubdate>
</bibl>

<bibl id="B7">
  <title><p>{The Mathematics of Infectious Diseases}</p></title>
  <aug>
    <au><snm>Hethcote</snm><fnm>HW</fnm></au>
  </aug>
  <source>SIAM REVIEW</source>
  <pubdate>2000</pubdate>
  <volume>42</volume>
  <issue>4</issue>
  <fpage>599</fpage>
  <lpage>-653</lpage>
</bibl>

<bibl id="B8">
  <title><p>Strong regularities in {World Wide Web} surfing</p></title>
  <aug>
    <au><snm>Huberman</snm><fnm>BA</fnm></au>
    <au><snm>Pirolli</snm><fnm>PLT</fnm></au>
    <au><snm>Pitkow</snm><fnm>JE</fnm></au>
    <au><snm>Lukose</snm><fnm>RM</fnm></au>
  </aug>
  <source>Science</source>
  <pubdate>1998</pubdate>
  <volume>280</volume>
  <fpage>95</fpage>
  <lpage>-97</lpage>
</bibl>

<bibl id="B9">
  <title><p>Using a Model of Social Dynamics to Predict Popularity of
  News</p></title>
  <aug>
    <au><snm>Lerman</snm><fnm>K</fnm></au>
    <au><snm>Hogg</snm><fnm>T</fnm></au>
  </aug>
  <source>Proc. of the 19th Intl. World Wide Web Conference (WWW2010)</source>
  <publisher>NY: ACM</publisher>
  <pubdate>2010</pubdate>
  <fpage>621</fpage>
  <lpage>-630</lpage>
</bibl>

<bibl id="B10">
  <title><p>Using Stochastic Models to Describe and Predict Social Dynamics of
  Web Users</p></title>
  <aug>
    <au><snm>Lerman</snm><fnm>K</fnm></au>
    <au><snm>Hogg</snm><fnm>T</fnm></au>
  </aug>
  <source>Submitted to ACM Transactions on Intelligent Systems and
  Technology</source>
  <pubdate>2011</pubdate>
</bibl>

<bibl id="B11">
  <title><p>The Double {Pareto}-Lognormal Distribution: A New Parametric Model
  for Size Distributions</p></title>
  <aug>
    <au><snm>Reed</snm><fnm>WJ</fnm></au>
    <au><snm>Jorgensen</snm><fnm>M</fnm></au>
  </aug>
  <source>Communications in Statistics: Theory and Methods</source>
  <pubdate>2004</pubdate>
  <volume>33</volume>
  <fpage>1733</fpage>
  <lpage>-1753</lpage>
</bibl>

<bibl id="B12">
  <title><p>Diversity of User Activity and Content Quality in Online
  Communities</p></title>
  <aug>
    <au><snm>Hogg</snm><fnm>T</fnm></au>
    <au><snm>Szabo</snm><fnm>G</fnm></au>
  </aug>
  <source>Proc. of the Third International Conference on Weblogs and Social
  Media (ICWSM2009)</source>
  <publisher>AAAI</publisher>
  <pubdate>2009</pubdate>
  <fpage>58</fpage>
  <lpage>-65</lpage>
</bibl>

<bibl id="B13">
  <title><p>On Fitting The {Poisson} Lognormal Distribution to
  Species-Abundance Data</p></title>
  <aug>
    <au><snm>Bulmer</snm><fnm>M. G.</fnm></au>
  </aug>
  <source>Biometrics</source>
  <pubdate>1974</pubdate>
  <volume>30</volume>
  <fpage>101</fpage>
  <lpage>-110</lpage>
</bibl>

<bibl id="B14">
  <title><p>Statistical Modelling of {Poisson}/Log-Normal Data</p></title>
  <aug>
    <au><snm>Miller</snm><fnm>G</fnm></au>
  </aug>
  <source>Radiation Protection Dosimetry</source>
  <pubdate>2007</pubdate>
  <volume>124</volume>
  <fpage>155</fpage>
  <lpage>-163</lpage>
</bibl>

<bibl id="B15">
  <title><p>Negative Binomial Regression</p></title>
  <aug>
    <au><snm>Hilbe</snm><fnm>JM</fnm></au>
  </aug>
  <publisher>Cambridge Univ. Press</publisher>
  <pubdate>2008</pubdate>
</bibl>

<bibl id="B16">
  <title><p>Bootstrap Methods: Another Look at the Jackknife</p></title>
  <aug>
    <au><snm>Efron</snm><fnm>B</fnm></au>
  </aug>
  <source>Annals of Statistics</source>
  <pubdate>1979</pubdate>
  <volume>7</volume>
  <fpage>1</fpage>
  <lpage>-26</lpage>
</bibl>

<bibl id="B17">
  <title><p>Power-law Distributions in Empirical Data</p></title>
  <aug>
    <au><snm>Clauset</snm><fnm>A</fnm></au>
    <au><snm>Shalizi</snm><fnm>CR</fnm></au>
    <au><snm>Newman</snm><fnm>M. E. J.</fnm></au>
  </aug>
  <source>SIAM Review</source>
  <pubdate>2009</pubdate>
  <volume>51</volume>
  <fpage>661</fpage>
  <lpage>-703</lpage>
</bibl>

<bibl id="B18">
  <title><p>Experimental Study of Inequality and Unpredictability in an
  Artificial Cultural Market</p></title>
  <aug>
    <au><snm>Salganik</snm><fnm>M.J.</fnm></au>
    <au><snm>Dodds</snm><fnm>P.S.</fnm></au>
    <au><snm>Watts</snm><fnm>D.J.</fnm></au>
  </aug>
  <source>Science</source>
  <pubdate>2006</pubdate>
  <volume>311</volume>
  <fpage>854</fpage>
</bibl>

<bibl id="B19">
  <title><p>Description and prediction of Slashdot activity</p></title>
  <aug>
    <au><snm>Kaltenbrunner</snm><fnm>A.</fnm></au>
    <au><snm>Gomez</snm><fnm>V.</fnm></au>
    <au><snm>Lopez</snm><fnm>V.</fnm></au>
  </aug>
  <source>Proc. 5th Latin American Web Congress (LA-WEB 2007)</source>
  <pubdate>2007</pubdate>
</bibl>

<bibl id="B20">
  <title><p>Analysis of Social Voting Patterns on {Digg}</p></title>
  <aug>
    <au><snm>Lerman</snm><fnm>K.</fnm></au>
    <au><snm>Galstyan</snm><fnm>A.</fnm></au>
  </aug>
  <source>Proceedings of the 1st ACM SIGCOMM Workshop on Online Social
  Networks</source>
  <pubdate>2008</pubdate>
</bibl>

<bibl id="B21">
  <title><p>Strong regularities in online peer production</p></title>
  <aug>
    <au><snm>Wilkinson</snm><fnm>DM</fnm></au>
  </aug>
  <source>EC '08: Proceedings of the 9th ACM conference on Electronic
  commerce</source>
  <publisher>New York, NY, USA: ACM</publisher>
  <pubdate>2008</pubdate>
  <fpage>302</fpage>
  <lpage>-309</lpage>
</bibl>

<bibl id="B22">
  <title><p>Novelty and collective attention</p></title>
  <aug>
    <au><snm>Wu</snm><fnm>F</fnm></au>
    <au><snm>Huberman</snm><fnm>BA</fnm></au>
  </aug>
  <source>Proceedings of the National Academy of Sciences</source>
  <pubdate>2007</pubdate>
  <volume>104</volume>
  <issue>45</issue>
  <fpage>17599</fpage>
  <lpage>-17601</lpage>
</bibl>

<bibl id="B23">
  <title><p>Viral, Quality, and Junk Videos on {YouTube}: Separating Content
  From Noise in an Information-Rich Environment</p></title>
  <aug>
    <au><snm>Crane</snm><fnm>R</fnm></au>
    <au><snm>Sornette</snm><fnm>D</fnm></au>
  </aug>
  <source>Proc. of the AAAI Symposium on Social Information Processing</source>
  <editor>K. Lerman and others</editor>
  <pubdate>2008</pubdate>
  <fpage>18</fpage>
  <lpage>-20</lpage>
</bibl>

<bibl id="B24">
  <title><p>Social influence and the diffusion of user-created
  content</p></title>
  <aug>
    <au><snm>Bakshy</snm><fnm>E</fnm></au>
    <au><snm>Karrer</snm><fnm>B</fnm></au>
    <au><snm>Adamic</snm><fnm>LA</fnm></au>
  </aug>
  <source>Proc. of the 10th ACM Conf. on Electronic Commerce (EC09)</source>
  <publisher>NY: ACM</publisher>
  <pubdate>2009</pubdate>
  <fpage>325</fpage>
  <lpage>-334</lpage>
</bibl>

<bibl id="B25">
  <title><p>{Early Warning Analysis for Social Diffusion Events}</p></title>
  <aug>
    <au><snm>Colbaugh</snm><fnm>R</fnm></au>
    <au><snm>Glass</snm><fnm>K</fnm></au>
  </aug>
  <source>Proceedings of IEEE International Conferences on Intelligence and
  Security Informatics</source>
  <pubdate>2010</pubdate>
</bibl>

<bibl id="B26">
  <title><p>Distinguishing influence-based contagion from homophily-driven
  diffusion in dynamic networks</p></title>
  <aug>
    <au><snm>Aral</snm><fnm>S</fnm></au>
    <au><snm>Muchnik</snm><fnm>L</fnm></au>
    <au><snm>Sundararajan</snm><fnm>A</fnm></au>
  </aug>
  <source>Proceedings of the National Academy of Sciences</source>
  <pubdate>2009</pubdate>
  <volume>106</volume>
  <issue>51</issue>
  <fpage>21544</fpage>
  <lpage>-21549</lpage>
  <url>http://dx.doi.org/10.1073/pnas.0908800106</url>
</bibl>

<bibl id="B27">
  <title><p>{GroupLens}: Applying Collaborative Filtering to {Usenet}
  News</p></title>
  <aug>
    <au><snm>Konstan</snm><fnm>J. A.</fnm></au>
    <au><snm>Miller</snm><fnm>B. N.</fnm></au>
    <au><snm>Maltz</snm><fnm>D.</fnm></au>
    <au><snm>Herlocker</snm><fnm>J. L.</fnm></au>
    <au><snm>Gordon</snm><fnm>L. R.</fnm></au>
    <au><snm>Riedl</snm><fnm>J.</fnm></au>
  </aug>
  <source>Communications of the ACM</source>
  <pubdate>1997</pubdate>
  <volume>40</volume>
  <issue>3</issue>
  <fpage>77</fpage>
  <lpage>-87</lpage>
  <url>citeseer.ist.psu.edu/konstan97grouplens.html</url>
</bibl>

</refgrp>
} 